\documentstyle[epsfig,12pt]{article}

\begin{document}
\topmargin 0pt
\oddsidemargin 5mm

\setcounter{page}{1}

\hspace{8cm}{Preprint YerPhI 1484(1)-97}
\vspace{2cm}
\begin{center}

{\large{ONE DIMENSIONAL NONLINEAR WAKE-FIELDS EXCITED}}\\
{\large{IN A COLD PLASMA BY CHARGED BUNCHES}} \\

{A.G. KHACHATRYAN}\\
\vspace{1cm}
{\em Yerevan Physics Institute}\\
{Alikhanian Brothers St. 2, Yerevan 375036, Republic of Armenia}
\end{center}

\vspace {5mm}
\centerline{{\bf{Abstract}}}

One dimensional nonlinear plasma wake-fields excited by a single bunch
and by series of bunches are considered. Essential differences are brought
to light between negatively charged bunch case and positively one. 
The bunches with nonuniform distributions of density are investigated. 
The obtained results shows dependence of excited potential electric fields on 
bunches parameters and allows to choose these parameters optimal.
\newpage

\section{Introduction}

\indent The electromagnetic waves, excited in plasma by charged bunches, 
can be 
used both for focusing of bunches and for charge acceleration \cite{A}. 
In the relativistic bunch case an amplitude of excited one dimensional
nonlinear plasma waves can essentially exceed the conventional 
wavebreaking field $E_{*}=m_ev_0\omega_p/e$ ($v_0$ is the phase velocity, 
$\omega_p=(4\pi n_0e^2/m_e)^{1/2}$ is the electron plasma frequency, $n_0$-
plasma electron density in equilibrium) and reaches of the value 
$\tilde{E}=[2(\gamma-1)]^{1/2}$, where $\gamma=(1-\beta^2)^{-1/2},
\beta=v_0/c$, and $\tilde{E}$ is normalized on $E_{*}$. Acceleration rate 
in the wake-fields can reaches of a few GeV/m, that is much more, than was 
reached on conventional accelerators.

The one dimensional nonlinear wake-fields theory was developed in series 
of works (see e.g. \cite{B}-\cite{H} and references therein). Maximum of 
the accelerated electrons energy, as it was shown in \cite{H}, can reach of 
value $4m_ec^2\gamma^3$.

In this paper bunches with the different parameters investigated 
analytically and by numerical simulation (including nonuniform and 
positively charged bunches).

\section{One Dimensional Nonlinear Wake-fields}

Consider a cold electron plasma with moving through the plasma, in $Z$
direction charged bunch (or train of bunches), which transverse sizes one
can consider infinite. The subject of our investigation is the steady 
wakefields excited by such bunches. Plasma ions assumed immobile because 
of their large mass ($m_e/m_i \ll 1$). As usually we also assume, that field 
in front of the bunch (or bunches) is absent, potential and strength of the
electric field are continuous in all space. Then, from the continuety 
equation, relativistic equation of motion for plasma electrons and the 
Poisson equation one can obtain following equation for steady fields 
\cite{E},\cite{H}
\begin{equation}
\label{1}
\frac{d^2F}{dz^2}+\beta^2\gamma^2\left(1-\frac{\beta 
F}{(F^2-\gamma^{-2})^{1/2}}\right)+\beta^2\alpha(z)=0,
\end{equation}
where $F=1+|e|\varphi/m_ec^2 \ge \gamma^{-1},\varphi$ is the electric 
potential, $z=k_p(Z-v_0t),k_p=\omega_p/v_0$ (here the phase 
velocity $v_0$ is equal to bunch velocity), $\alpha(z)=(q/|e|)n_b(z)/n_0,e$
is the electron charge, $q$ is the charge of the bunch particles, $n_b(z)$ is
their density. Normalized on $E_{*}$ the strength of electric field obeys the
formula $E_z(z) \equiv E=-(1/\beta^2)dF/dz$. The plasma electrons 
dimensionless velocity as function of $F$ is 
\begin{equation}
\label{2}
\beta_e=[\beta-(F^2-\gamma^{-2})^{1/2}]/(\beta^2+F^2).
\end{equation}

Inside uniform bunches and outside of bunches  Eq. (\ref{1})  can be 
rewritten in the form \cite{E}
\begin{eqnarray}
\label{3}
\frac{d^2F}{dz^2}+\frac{dU}{dF}=0,\nonumber \\
U=\beta^2[(\gamma^2+\alpha)F-\beta\gamma^2(F^2-\gamma^{-2})^{1/2}].
\end{eqnarray}
Formally Eq. (\ref{3}) describes one-dimensional motion of a particle
in a field with potential $U(F)$. Analysis of the function $U(F)$ allows 
to determine both qualitative behaviour of the field and some quantitative 
values (such as electric field amplitude in the bunch, the wave-breaking 
field, maximum value of electric potential and other). Moreover, it can be 
obtained analytical solution of Eq. (\ref{3}), which includes the 
elliptic functions \cite{C},\cite{E}-\cite{G}. However, even in the case of 
uniform bunch, the analytical description of the field behind bunch in 
general case is practically impossible. For nonuniform bunches the problem 
analytically is not solved yet.

Outside the bunch (where $\alpha(z)=0$) "potential" $U(F)$ has one 
minimum, therefore the field is periodical and for fixed $\gamma$ fully determines 
by electric field amplitude $E_{mp} \le \tilde{E}$. It is necessary to 
note that Eq. (\ref{3}) with $\alpha=0$ describes also the wake 
wave, excited by other way (for example, by laser pulse).

Nonlinear wake wavelength $\Lambda_p$ increases as the amplitude $E_{mp}$
increases. When $\gamma \gg 1$ the dependence $\Lambda_p(E_{mp})$ practically
not depend on $\gamma$ and is presented on Figure 1 (notice, that according
to accepted in (\ref{1}) variables, value $\Lambda_p=2\pi$ corresponds to 
the linear approach). Near by the "breaking" $(E_{mp} \approx \tilde{E})$
the wake wavelength in the case of $\gamma \gg 1$ is approximately equal 
to $4(2\gamma)^{1/2}$ \cite{C}, \cite{F} and maximum value of the 
dimensionless electric potential is $F_{mp} \approx (1+\beta^2)\gamma$  
\cite{E}. In general case $F_{mp}$ increases as $E_{mp}$ increases (see 
Fig.1).

The energy gain of electrons (or positrons) accelerating in the wake wave can
reach of value $4m_ec^2\gamma^3$ \cite{E}. Really, the relativistic 
equation of motion of accelerating electron in the frame of reference 
moving with the wave velocity is
\begin{equation}
\label{4}
d(\beta'_a\gamma'_a)/dt'=cdF'/dZ'.
\end{equation}
Taking into account, that $dZ'=c\beta'_adt'$ and 
$\beta'_ad(\beta'_a\gamma'_a)=d\gamma'_a$ from (\ref{4}) follows 
$\gamma'_a=\gamma'_a(0)+F'-F'(0),$ where $\gamma'_a(0)$ and $F'(0)$ are 
initial values. In the laboratory frame of reference $\gamma_a$ one can 
obtain using known relativistic transformations $\varphi=\gamma\varphi'$
and $\gamma_a=\gamma\gamma'_a(1+\beta\beta'_a$). In the case $\gamma \gg 1$
and $\beta_a(0) \ge \beta$, choosing $F(0)\approx 1/\gamma$ and $F\approx 
F_{mp} \approx 2\gamma$ (that corresponds to $E_{mp} \approx \tilde{E}$)
obtain $\gamma_a \approx 4\gamma^3.$ In this case acceleration occur on 
the half of the wavelength;the acceleration length is proportional to 
$\gamma^{5/2}$ \cite{E}. For example, when $n_0=3\cdot 10^{12}cm^{-3}$ and 
$\gamma=10$, the acceleration length is $l_a \approx 1m,\Lambda_p \approx 
5cm$
and $(\gamma_a)_{max} \approx 4000$. The result $(\gamma_a)_{max} \sim 
\gamma^3$ is consequence both of correlations $E_{mp} \sim \gamma^{1/2},
\Lambda_p \sim \gamma^{1/2}$ and relativistically large mass of the 
accelerating particle (i.e. $\beta_a \approx \beta \approx 1$), that causes 
long keeping back of the particle in accelerating phase. For arbitrary 
wake wave amplitude the maximum value $F_{mp}$ can be obtained from Fig.1.

For the large $\gamma$ the acceleration length can surpasses the 
laboratory plasma sizes. In the cosmos (for example, in the blanket of 
supernova), where the plasma can be considered unlimited, acceleration in 
the strong plasma fields described here can originate high energy cosmic 
rays.

\section{Single Bunch}

Consider a single uniform bunch:$n_b=const$, when $-d \le z \le 0$ and 
$n_b=0$ elsewhere. The field inside the uniform bunch describes by Eq. 
(\ref{3}). "Potential" $U(F)$ is maximum when $F=1/\gamma$ and is minimum 
when \begin{equation}
\label{5}
F=F_c=|\gamma^2+\alpha|/(\gamma 
\Gamma^{1/2}),\Gamma=\gamma^2(1+2\alpha)+\alpha^2.
\end{equation}
From (\ref{5}) follows, that in the case $-1/(1+\beta) \equiv \alpha_1 
<\alpha<\infty$ function $U(F)$ has minimum and the field inside bunch 
is periodical;$F$ changes in the limits
\begin{equation}
\label{6}
1/\gamma \le F \le [(\gamma^2+\alpha)^2+\beta^2\gamma^4]/\gamma\Gamma.
\end{equation}
When $\alpha=\alpha_1,U(F)$ monotonously decreases as $F$ increases and 
tends to zero when $F \rightarrow \infty$. In this case the electric field 
strength inside the bunch grows from zero at the bunch head ($z=0$) and 
tends to some constant value (see (\ref{7})). When $\alpha <\alpha_1,U(F)$
decreases unlimitedly, $E$ grows from the bunch head to its tail 
($z=-d$), and for bunch length $d$ more than some value $d_{max}$ the wake 
behind the bunch "breaks".

When $U(F(z=0)=1)>U_{max}=U(F=1/\gamma)$ (that takes place, when 
$\alpha>\gamma$), in enough long bunch $F$ becomes less than 
$1/\gamma$, that is
using in this paper cold fluid approach becomes inapplicable. Thus, in the 
cases $-\infty<\alpha<\alpha_1$ and $\alpha>\gamma$ the uniform bunch 
length can not exceed some value $d_{max}$;the wake amplitude in the case 
of $d=d_{max}$ is equal to $\tilde{E}$. Dependence of the $d_{max}$ on 
$\alpha$ for $\gamma=10$ presented on Figure 2.

In the case of $\alpha_1 \le \alpha \le \gamma$ the uniform bunch length 
can be arbitrary. The field inside the bunch is periodical and has 
amplitude \cite{E}, \cite{F}
\begin{equation}
\label{7}
E_{mb}=[2(1+\alpha-\Gamma^{1/2}/\gamma)]^{1/2}/\beta.
\end{equation}
Dependence of the wavelength inside the bunch $\Lambda_b$ on $\alpha$ for
$\gamma \gg 1$ shown on Figure 3.

In the linear approach $E_{mp} \sim \alpha \quad (|\alpha| \ll 1)$ and 
differences between the electron bunch case and the positron one reduce 
to changing of electric field sign (saying electron or positron bunch we 
mean negatively or positively charged bunch, respectively). In nonlinear 
theory these differences are essential. The wavelength $\Lambda_b$ in the 
electron bunch increases as $|\alpha|$ increases when $\alpha_1<\alpha<0$
(see Fig.3), and in the case $\gamma \gg 1$ and $\alpha \approx -0.5$ is 
nearly equal to $8\gamma$ \cite{C}, \cite{E}, \cite{F}, that is much more
than linear plasma wavelength. In the positron bunch $\Lambda_b$ decreases 
as $\alpha$ increases. Note, that difference of the nonlinear wavelength 
from linear one, in the case $\gamma \gg 1$ is palpable even for $|\alpha|
\ll 1$. From (\ref{3}), (\ref{5}) and the boundary conditions one can see, 
that $F(z) \ge 1$ in the electron bunch, and $F(z) \le 1$ in the positron 
bunch. That is, according to (\ref{2}), $\beta_e(z) \le 0$ in the electron 
bunch and $\beta_e(z) \ge 0$ in the positron one.

Equation(\ref{1}) was investigated numerically for uniform, parabolic and 
linear charge density distributions in the bunch
\begin{eqnarray}
\label{8}
\alpha_u=\alpha_{u0}, \nonumber \\
\alpha_p=\alpha_{p0}[1-(z-d/2)^2]/(d/2)^2, \nonumber \\
\alpha_l=\alpha_{l0}|z|/d, \nonumber \\
-d \le z \le 0.
\end{eqnarray}
Bunches with profiles (\ref{8}), but with the same length and same total 
charge are compared. That means $|\alpha_{p0}|=(3/2)|\alpha_{u0}|,
|\alpha_{l0}|=2|\alpha_{u0}|$.

Typical dependence of the wake wave amplitude $E_{mp}$ on bunch length 
for distributions (\ref{8}), in the case of periodical field inside the 
uniform bunch (i.e. when $\alpha_1<\alpha_{u0}<\gamma$) shown on Figure 4.
For the uniform bunch this dependence is periodical (on Fig. 4 shown one 
period only). The wake wave amplitude behind the uniform bunch is maximum 
and equal to $E_{mp}=2|\alpha_{u0}|\gamma/\Gamma^{1/2}$ \cite{E}, \cite{F}
when $d=(n+1/2)\Lambda_b$ ($n$ is integer) and equal to zero when 
$d=n\Lambda_b$. For the parabolic and linear profiles choice $d \approx 
(3/4)\Lambda_b$ is optimal. One can see also, that in the case $d>\Lambda_b$
the "parabolic" bunches becomes less effective and the wake wave 
amplitude behind the bunch with linear profile weakly depends on $d$. For
a short bunch ($d<\Lambda_b/2$) $E_{mp}(d)$ almost not depends on the bunch
profile.

\section{Trains of Bunches}

The results presented above demonstrates, that for the strong wake wave 
(with $E_{mp} \sim \tilde{E}$) excitation it is necessary relativistic 
either negatively charged bunch with $\alpha \le -0.5$ or positively 
charged bunch with $\alpha \ge \gamma$. However, on experiments density of 
bunches is often much less of the plasma density, i.e. $|\alpha| \ll 1$. 
Therefore, it is naturally to try excite the strong waves by series of 
bunches. For the case of uniform bunches with the same densities this 
problem was investigated analytically in \cite{G}. The wake wave amplitude 
behind the train of uniform bunches is maximum when 
$d_i=(n_i+1/2)\Lambda_{bi}$ and $l_j=(n_j+1/2)\Lambda_{pj}$, where $d_i$ is
the length of $i$-th bunch, $l_j$ is the spacing between $j$-th and 
$(j+1)$-th
bunches, $\Lambda_{bi}$ is the wavelength in $i$-th bunch, $\Lambda_{pj}$ is 
the wavelength behind $j$-th bunch, $n_i$ and 
$n_j$ are integer. The wake wave amplitude behind such optimized train 
depending on number of bunch $N$ is \cite{G}:
\begin{equation}
\label{9}
E_{mp}(N) \approx (2/\beta)sh(|\alpha|\beta N),|\alpha| \ll 1.
\end{equation}
From (\ref{9}) follows, that the wave amplitude quickly grows as $N$ 
increases and reaches the value $\tilde{E}$ when
\begin{equation}
\label{10}
N \approx (|\alpha|\beta)^{-1}\ln[(\gamma+1)^{1/2}/2+(\gamma-1)^{1/2}/2].
\end{equation}
The electric field in the optimised train, obtained by numerical 
simulation of equation (\ref{1}) presented on Figure 5. The numerical 
results showed high exactness of the expressions (\ref{9}) and (\ref{10}).

Often in experiments using trains of bunches with $d_i=d=const,l_j=l=const$
and $|\alpha(z)| \ll 1$. Analytical discription of the fields excited in 
such trains is very difficult. The numerical simulation showed, that for 
profiles (\ref{8}) in the case of $d$ and $l$ optimal for first bunch
(see Sec. 3) the wake wave amplitude behind the train grows almost 
linearly up to $N=N_{*} \sim 0.5/|\alpha_{u0}|. E_{mp}(N_{*})$ increases
as $|\alpha_{u0}|$ and (or) $\gamma$ increases. When the wave becomes 
essentially nonlinear (when $N \ge N_{*}$) coherence of the waves 
exciting by the bunches  breaking because of increasing of the wave 
length and the wave amplitude behind train decreases almost up to zero as
$N$ increases. Dependence $E_{mp}(N)$ for $N \gg N_{*}$ is nearly periodical.

It was ascertained, that trains of short bunches ($d \ll \Lambda_b/2$ and 
$l \sim \Lambda_p/2$) are not effective as in this case $E_{mp}(N) \le 
E_{mp}(1)$. When $d \ll \Lambda_b$ and $l \ll \Lambda_p$ the train 
behaves as one bunch with averaged along the train length charge density.

\section{Conclusions}

The analytical and numerical results presented above throws light upon a 
number of questions of the theory of strong plasma waves excited by charged
bunches. This results allow to choose parameters of the bunches optimal on 
future experiments.

The peculiarity of wake waves excitation by positively charged bunches 
are revealed. In particular, it was discovered, that the nonlinear 
wavelength in uniform positively charged bunch decreases as charge 
density increases.

For the excitation of wake wave with $E_{mp} \sim \tilde{E}$ more 
suitable is a bunch, consisting of heavy particles (for 
example, protons), as light particles (electrons, positrons) lose 
considerable part of their energy on length comparable with the bunch 
length. 
This is easy to show by comparing of the bunch energy with the energy of 
wake wave.

Note, that the results presented in this paper are also suitable for the 
case of two-dimensional cylindrical bunches with radius $a$ when 
$(\gamma c/\omega_pa)^2 \ll 1$ and $r^2 \ll a^2$ \cite{I}.

\section{Acknowledgments}

The author would like to thank A.Ts. Amatuni, S.S. Elbakian and E.V. 
Sekhpossian (Yerevan Physics Institute) for helpful discussions  and
G.A. Amatuni for the help in preparing the manuscript for publication.

This work was supported by the International Science and Technology 
Center, Project A-13.

\newpage
\epsfig{file=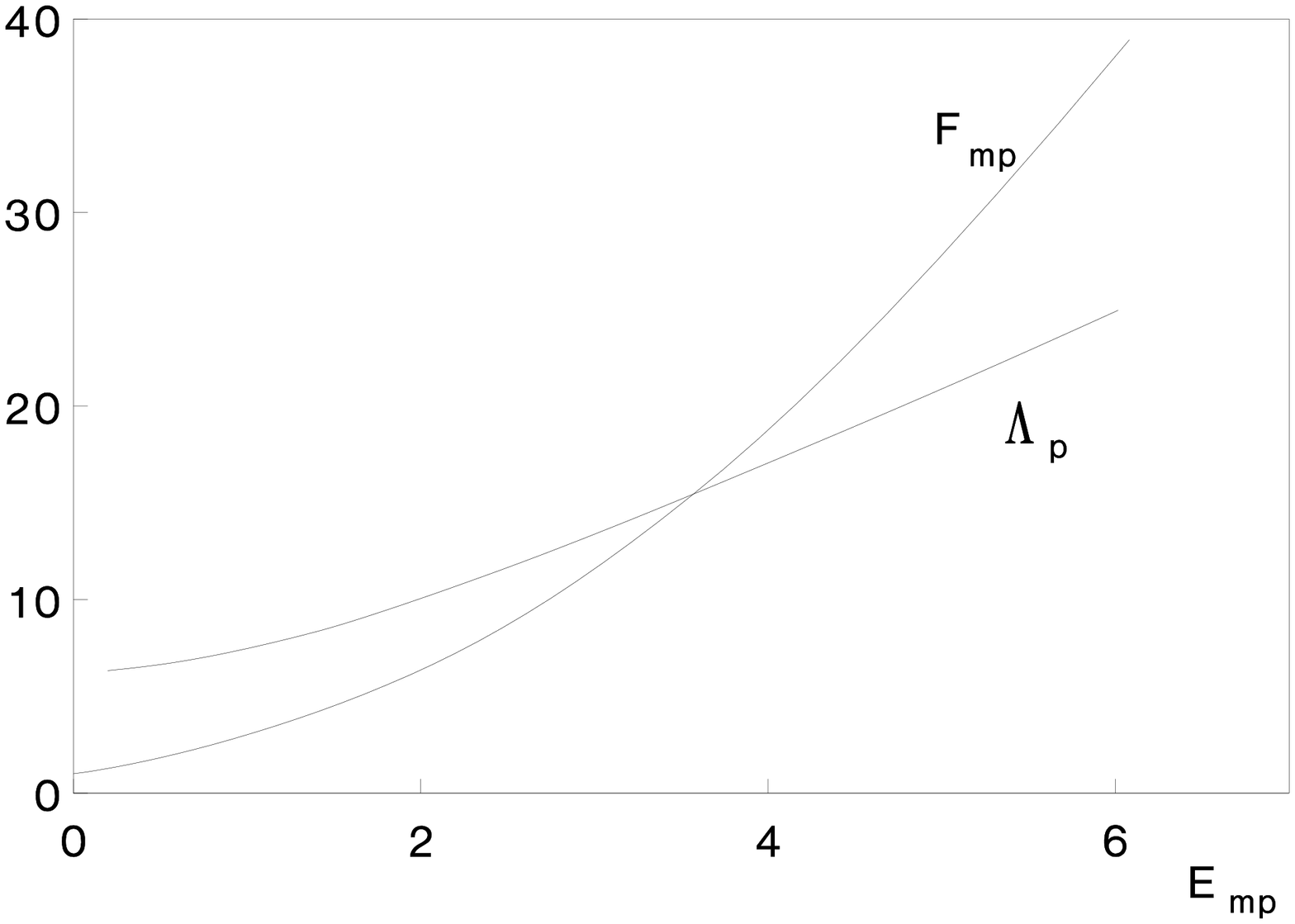,width=15cm,height=20cm}

Figure 1. Dependence of the wake wavelength and maximum value of the electric
potential on wake wave amplitude $E_{mp}$ (in the dimensionless units);
$\gamma \gg 1$.

\newpage
\epsfig{file=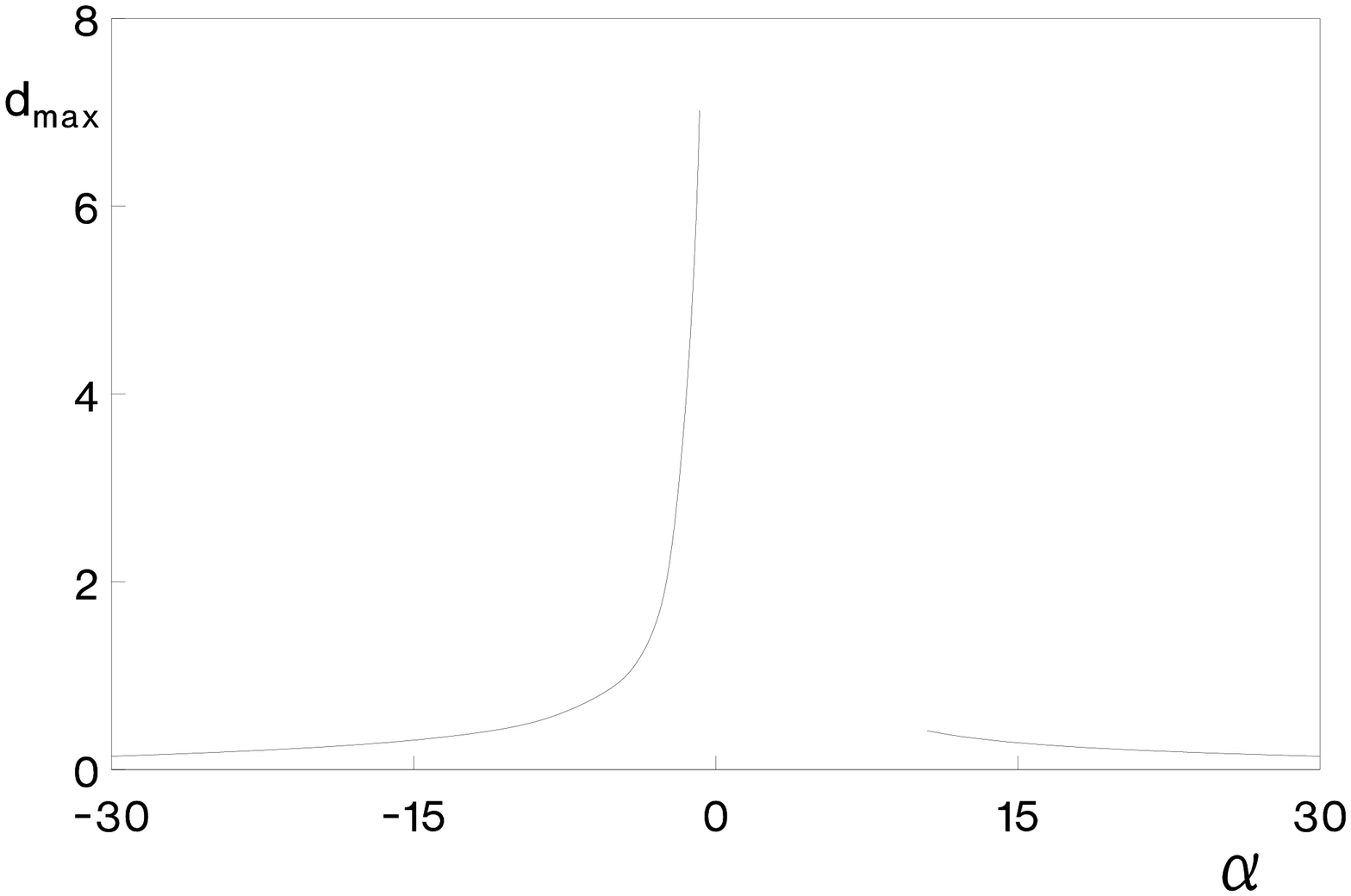,width=15cm,height=20cm}

Figure 2. The maximum permissible bunch length $d_{max}$ 
depending on the bunch charge density ($\gamma=10$).
\newpage 
\epsfig{file=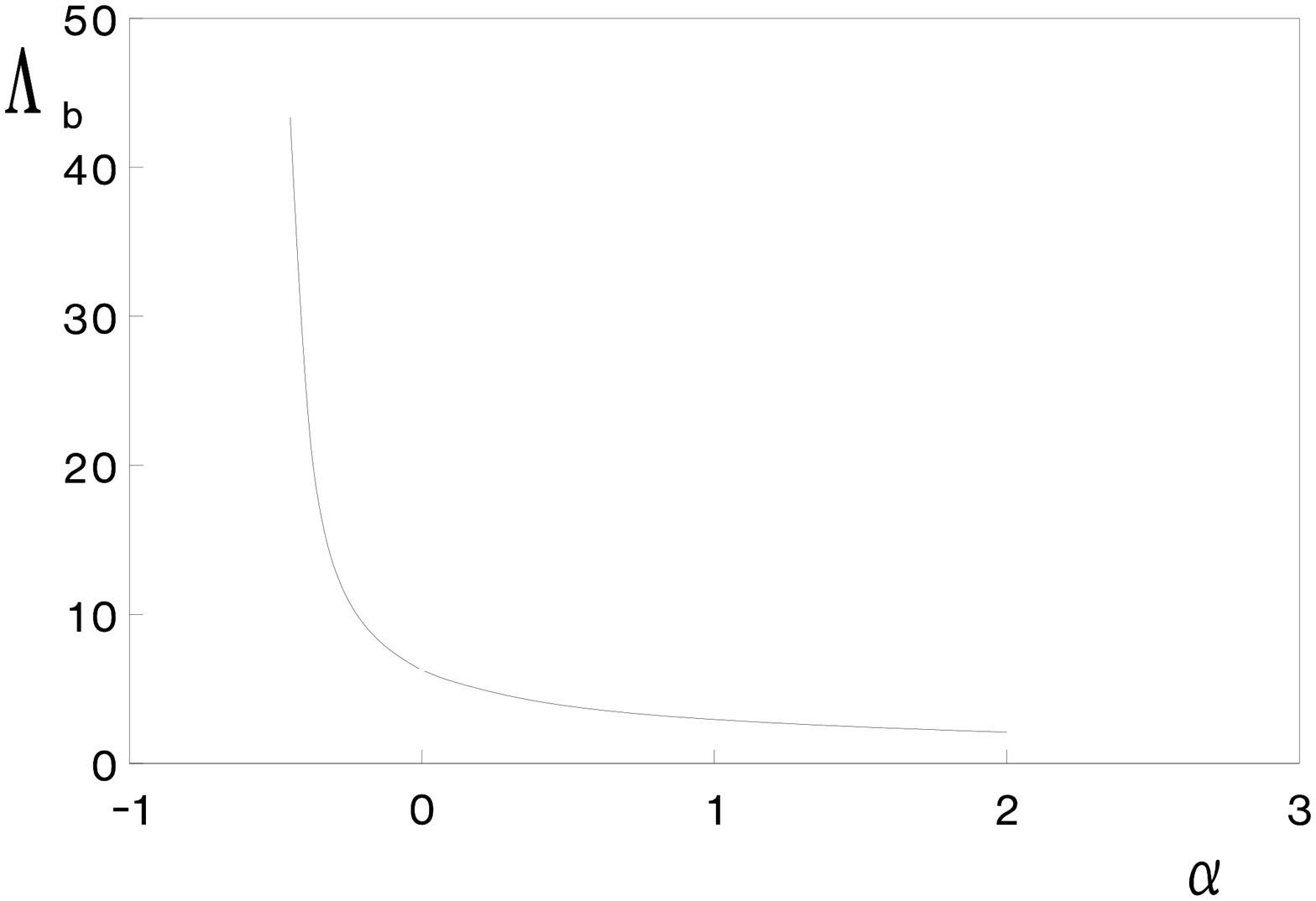,width=15cm,height=20cm}

Figure 3. Dependence of the wavelength in uniform bunch on the charge 
density.

\newpage
\epsfig{file=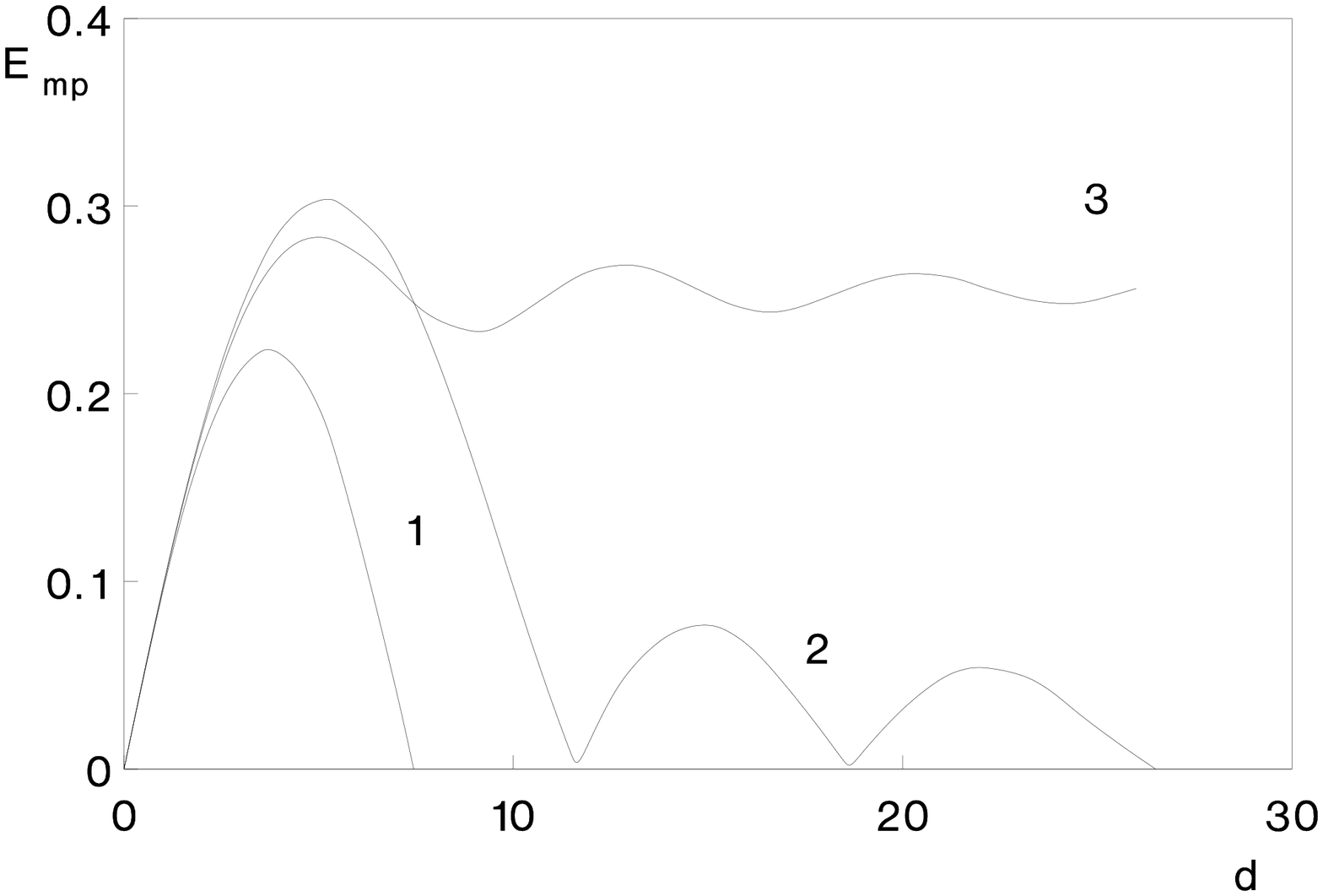,width=15cm,height=18cm}

Figure 4. The wake wave amplitude behind bunch depending on the bunch length
($\alpha_{u0}=0.1,\gamma=10$). 1-uniform bunch, 2-parabolic profile of 
the bunch charge density, 3-linear 
profile case.
\newpage

\epsfig{file=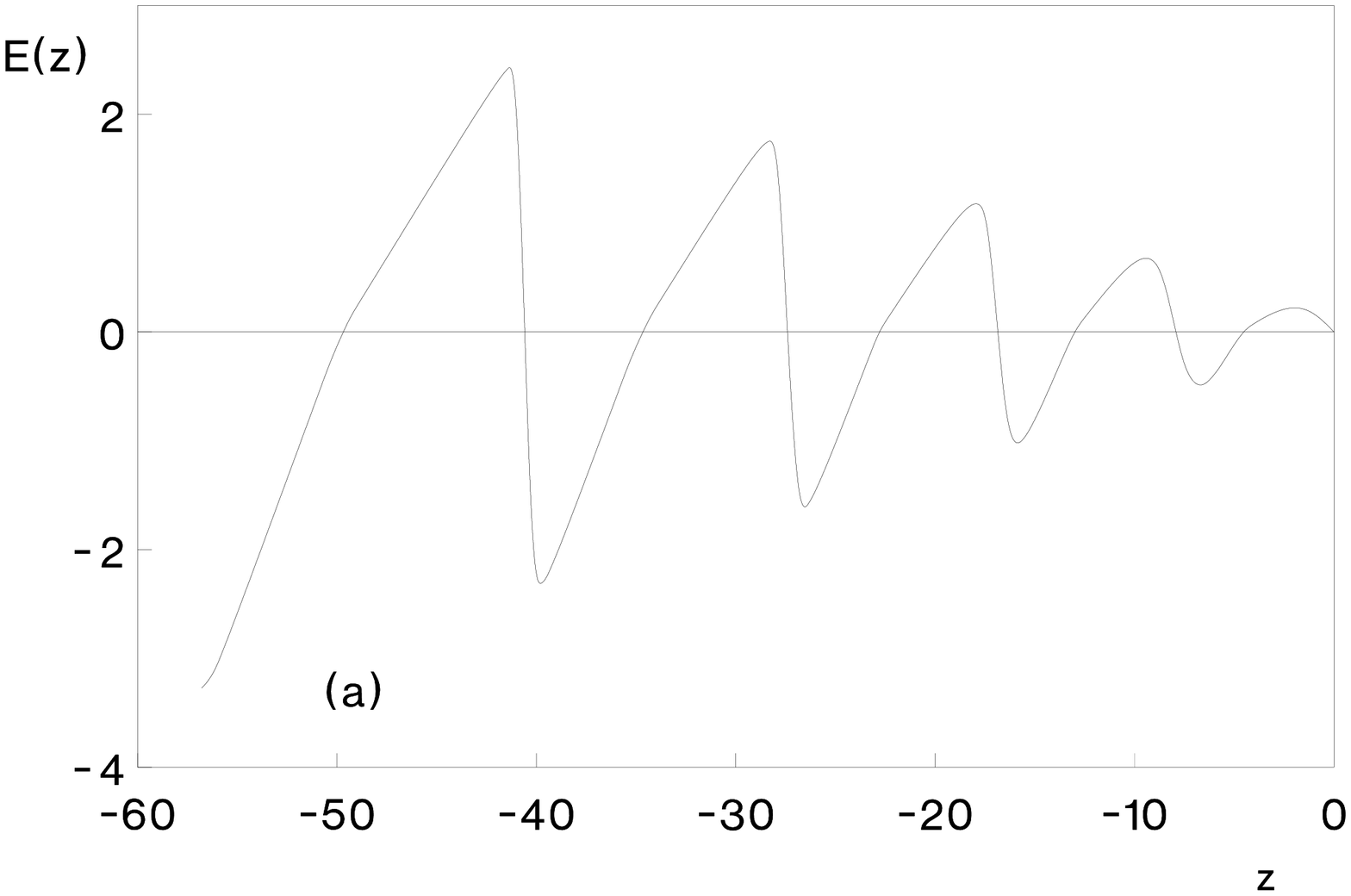,width=7cm,height=15cm}
\epsfig{file=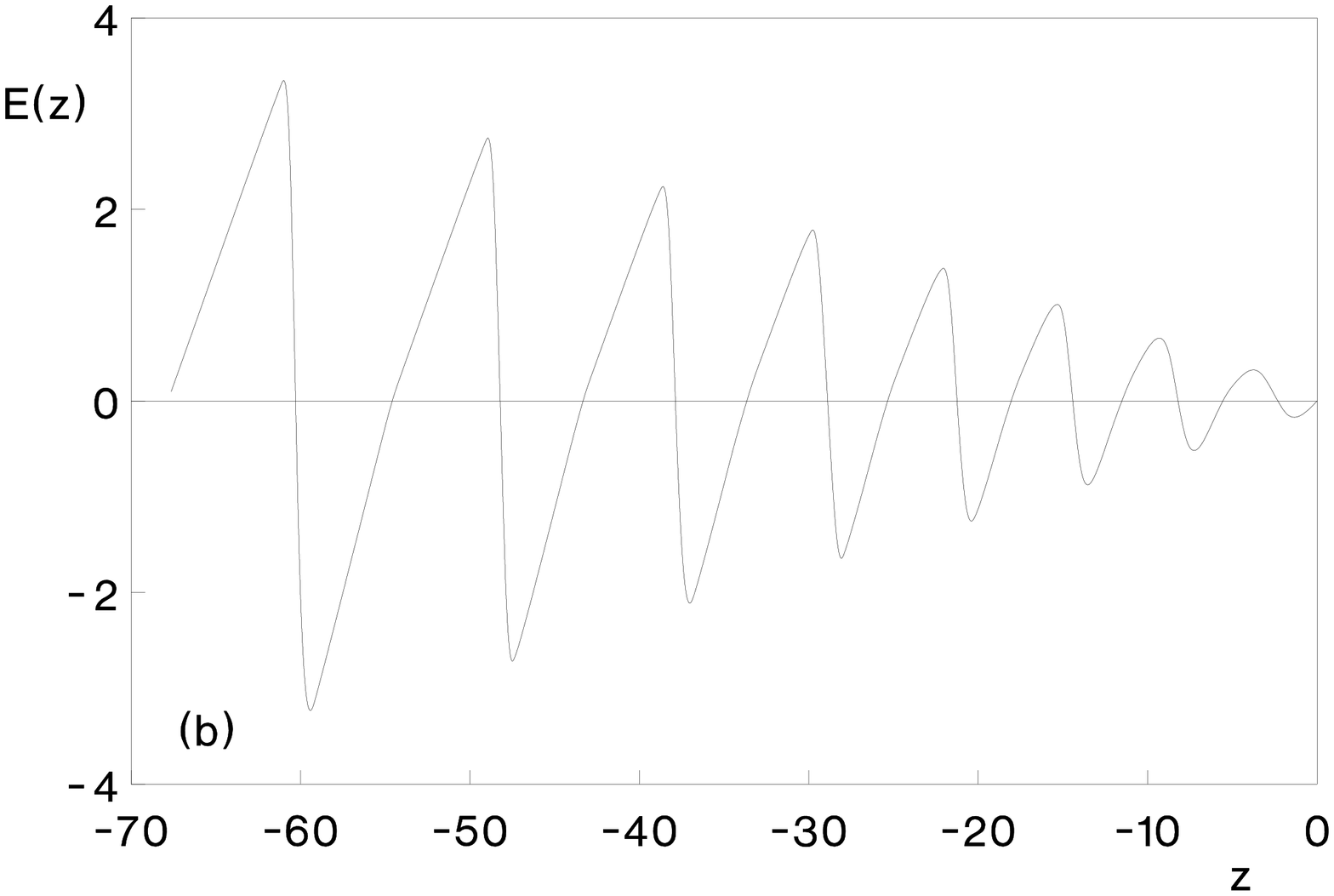,width=7cm,height=15cm}
 
Figure 5. The electric field in optimized train of uniform bunches.\\
 a) The train of negatively charged bunches ($\alpha=-0.2, \gamma=10$).
Inside bunches $E > 0$. Negative E corresponds to space between bunches.\\
 b) The train of positively charged bunches ($\alpha=0.2,\gamma=10$).
Inside bunches $E < 0$.Positive E corresponds to space between bunches.
\end{document}